\documentclass[aps,superscriptaddress,amsmath,amssymb,twocolumn,amsfonts]{revtex4-1}
\usepackage{amsmath}
\usepackage{graphicx}
\usepackage{hyperref}
\usepackage{cleveref}
\usepackage{color}
\usepackage{float}
\usepackage{soul}
\usepackage{ulem}
\usepackage{braket}

\newcommand\bea{\begin{eqnarray}}
\newcommand\eea{\end{eqnarray}}
\newcommand\beq{\begin{equation}}  
\newcommand\eeq{\end{equation}}

\newcommand{\pdag}{{\phantom\dagger}}

\begin{document}
\title{Theoretical analysis of multi-magnon excitations in resonant inelastic x-ray scattering spectra of two-dimensional antiferromagnets} 
\author{Subhajyoti Pal}
\affiliation{School of Physical Sciences, National Institute of Science Education and Research, a CI of Homi Bhabha National Institute, Jatni 752050, India}
\author{Umesh Kumar}
\affiliation{Department of Physics and Astronomy, Rutgers University, Piscataway, NJ 08854, USA}
\author{Prabhakar }
\affiliation{School of Physical Sciences, National Institute of Science Education and Research, a CI of Homi Bhabha National Institute, Jatni 752050, India}
\author{Anamitra Mukherjee}
\affiliation{School of Physical Sciences, National Institute of Science Education and Research, a CI of Homi Bhabha National Institute, Jatni 752050, India}

\begin{abstract}

Resonant inelastic x-ray spectroscopy (RIXS) has emerged as an important tool to explore magnetism in two-dimensional (2D) antiferromagnet realized in strongly correlated materials. Here we consider the Heisenberg model with nearest and next nearest neighbor hopping relevant to the study of magnetic excitations of the cuprate family.
We compute the RIXS cross-section within the ultra-short core-hole lifetime (UCL) expansion of the Kramers-Heisenberg scattering amplitude that allows perturbative solution within linear spin wave theory (LSWT). We report detailed results for both spin-conserving and  non-conserving channels. Apart from the widely discussed single magnon and bimagnon contributions, we show that three-magnon contributions in the spin non-conserving channel are useful to explain certain features of the RIXS data for two-dimensional cuprates. We confirm the qualitative correctness of the LSWT conclusions for the three-magnon excitation with exact diagonalization. Our work puts constraints on the dispersion  of the three-magnon in the Brillouin zone, opening new avenues for realizing higher modes of quasiparticles using RIXS.

\end{abstract}

\maketitle

\section{Introduction}
Strongly correlated materials are known to host exotic quantum phenomena such as superconductivity, strange metallicity, confinement, and deconfinement. Two-dimensional (2D) cuprates are known to host a number of such properties and are one of the most studied materials in the condensed matter community~\cite{RevModPhys.66.763, SCALAPINO1995329}. More recently, superconductivity has also been discovered in the infinite layer (IL) nickelates, which is proposed as a cuprate analogue~\cite{Li2019}. Superconductivity in 2D iridates has also been proposed due to their similarity to cuprates but never realized~\cite{doi:10.1146/annurev-conmatphys-031218-013113}. The magnetism in these 2D materials is understood to play a central role in these exotic phenomena~\cite{PWAnderson1987,HLu2021,PhysRevLett.108.177003}.

Resonant inelastic x-ray scattering has become an important tool in the recent past to explore magnetism in such strongly correlated materials~\cite{RevModPhys.93.035001, RevModPhys.83.705}. 
The Kramers-Heisenberg (KH) formalism employed to simulate the RIXS cross-section is complex and makes the problem challenging in interpreting RIXS data~\cite{RevModPhys.83.705, PhysRevB.85.064423, Nocera2018, Schlappa2018, PhysRevLett.112.147401, PhysRevB.88.195138}. Significant progress has been made in exploring quantum magnets using RIXS after the realization that it can allow for single spin-flip excitations at $L$-edge in cuprates~\cite{PhysRevLett.103.117003, PhysRevLett.104.077002, MorettiSala2011}
The ultra-short core-hole life expansion of RIXS maps the cross-section to spin-conserving (SC) and non-spin-conserving (NSC) channels at the Cu $L$-edge~\cite{UKumar2022, PhysRevLett.112.147401, PhysRevX.6.021020, PhysRevB.85.064421, PhysRevB.83.245133}. On the experimental side, progress has been made to resolve the polarization of the scattered RIXS spectra, therefore allowing for the filtering of spin-flip excitations~\cite{doi:10.1063/1.4900959, PhysRevB.99.134517}. 
Significant progress has been made in exploring the higher order corrections in the SC channel RIXS, as this channel dominates the edge that doesn't allow for single spin flips~\cite{PhysRevB.77.134428, Schlappa2018, UKumar2018}. However, there is limited work in exploring the higher order corrections in the NSC channel. The SC channel has been successful in identifying four-spinons in 1D cuprates~\cite{Schlappa2018, UKumar2018, UKumar2020}, multi-triplons in spin ladder~\cite{Schmiedinghoff2022, UKumar2019, Tseng2022} and multi-magnons in 2D cuprates~\cite{PhysRevB.103.224427}, in the phase space beyond traditional probes such as inelastic neutron scattering (INS) as this channel is usually inaccessible. On the other hand, the NSC channel of RIXS is usually understood to be equivalent to the INS probe, and the higher-order contributions have not been explored in the literature for the 2D cuprates. But, recently, a number of experiments have reported multi-magnons, attributing as higher order multi-magnons in the RIXS cross-section~\cite{PhysRevB.103.224427, PhysRevB.103.L140409}. 

Exploring the contributions of higher-order excitations has also greatly interested the INS community. For example, four-spinon contributions were reported in the INS response of 1D cuprate~\cite{Mourigal2013, PhysRevLett.111.137205}, multi-triplons in spin ladder cuprates~\cite{PhysRevLett.98.027403,PhysRevLett.87.127002}. The features beyond the conventional magnons were interpreted as fractionalized magnon excitations in 2D cuprates~\cite{DallaPiazza2015}. 

There has been a renewed interest in exploring the RIXS cross-section for the 2D cuprates~\cite{CLuo2014, Peng2017, MHe2020, PhysRevB.103.224427, PhysRevX.12.021041}. Recently, RIXS has been used to characterize the influence of apical oxygen in different cuprates ~\cite{Peng2017} by characterizing the features at higher energies at $(\pi, 0)$ and were characterized using further nearest neighbors.  
The anomalous feature in 2D cuprates has been identified as fractional spin excitations~\cite{PhysRevX.12.021041}. In the past, some of the features in the inelastic neutron spectra have been attributed to fractionalized spin excitations arising from magnon-magnon interactions~\cite{DallaPiazza2015}. This leads to complications in interpreting the additional features observed in RIXS experiments. Motivated by the recent identification of various correlation functions for the higher-order contributions in the UCL expansion of the KH formalism in 1D cuprates~\cite{UKumar2022}, we here explore these correlation functions in the context of 2D Heisenberg antiferromagnetic, realized in 2D cuprates \cite{PhysRevB.103.224427} and IL nickelates~\cite{HLu2021}. 

In this work, we report the distinct magnetic excitations realized in the SC and NSC channels of the UCL expansion of the RIXS cross-section employing standard linear spin-wave theory (LSWT) for 2D antiferromagnetic (AFM) Heisenberg Hamiltonian. Among the many interesting spin excitations, in particular, we find a remarkable quantitative match of our three-magnon spectrum over the entire Brillouin zone with purported multi-magnon excitation seen for La$_2$CuO$_4$\cite{PhysRevB.103.224427}. We further validate these conclusions with exact diagonalization.

Our paper is organized as follows. In sec.~\ref{sec:Methods}, we introduce the spin model for the 2D Heisenberg antiferromagnets explored in our work and present the mapping of the spin Hamiltonian to magnons. We also introduce the response functions for the corrections in UCL expansion of the RIXS cross-section explored in this work. In sec.~\ref{sec:ResultsAndDiscussion}, we present and discuss the results for the NSC and SC channel of RIXS cross-section evaluated using LSWT. We also compare the NSC channel LSWT results with exact-diagonalization for consistency. Finally, in sec.~\ref{sec:Conclusion}, we conclude our findings and comment on the recent Cu $L$-edge data of 2D cuprate reported in the literature.


\section{Method}\label{sec:Methods}
The two-dimensional (2D) antiferromagnets realized in 2D cuprates consist of the CuO$_2$ plaquettes. The corresponding one-band Hubbard model can be mapped to the spin half Heisenberg model and can very well capture the low energy spin dynamics realized in materials~\cite{UKumar2021, Nocera2018, Li2021, PhysRevB.37.9753}. We, therefore, here, start with the 2D spin model given by 
\begin{equation}
H_0 =J_{1} \sum_{\langle ij\rangle}{\bf S}_i \cdot {\bf S}_{j} + J_{2}\sum_{\langle\langle ij\rangle\rangle} {\bf S}_i \cdot {\bf S}_{j}. 
\label{eq:hamiltonian0}
\end{equation}
Here, $\langle\cdots\rangle$ and $\langle\langle\cdots\rangle\rangle$ indicate the sum over nearest neighbors (NN) and next-nearest neighbors (NNN) sites, respectively. $J_1$ and $J_2$ are the superexchange couplings between NN and NNN sites. ${\bf S}_i$ is the spin at site-$i$. We assume the z-axis as the quantization axis for the AFM. and consider a bipartite lattice for the AFMs, with sublattices-A and B. 
In this study, we consider AFM NN exchange. While the AFM NNN exchange is well-established for the cuprate family, from a theoretical standpoint, we also consider ferromagnetic (FM) NNN exchanges in our study for completeness.


\textit{Linear Spin Wave Theory for $H_0$:} We map the spin Hamiltonian to magnons using LSWT, using the usual notion of the bipartite lattice with sub-lattices; A and B. In the antiferromagnetic ground state, the Hamiltonian can be bosonized in LSWT, for which introduce the standard Holstein-Primakoff (HP) transformation as follows: 
\begin{equation}
\text{For sub lattice A} =
\begin{cases}
\hat{S}_i^+=\sqrt{2S} \sqrt{1-\frac{a^\dag_i a_i}{2S} }a_i \\
\hat{S}_i^-=\sqrt{2S} a^\dag_i \sqrt{1-\frac{a^\dag_i a_i}{2S} }\\
\hat{S}^z_i=S-a^\dag_i a_i
\end{cases} \label{eq:holstine_a}
\end{equation}

\begin{equation}
\text{For sub lattice B} =
\begin{cases}
\hat{S}_i^+=\sqrt{2S} b^\dag_i \sqrt{1-\frac{b^\dag_i b_i}{2S} }\\
\hat{S}_i^-=\sqrt{2S}\sqrt{1-\frac{b^\dag_i b_i}{2S} } b_i \\
\hat{S}^z_i=-S+b^\dag_i b_i
\end{cases} \label{eq:holstine_b}
\end{equation}

To diagonalize the Hamiltonian $\hat{H}_0$ in Eq: \ref{eq:hamiltonian0}, we need to introduce the Bogoliubov transformation, which in reciprocal space is defined as: 

\begin{equation}
\begin{bmatrix}
\alpha_{\mathbf{k}} &\\
\beta_{-\mathbf{k}}^\dag
\end{bmatrix} = \begin{bmatrix}
u_{\mathbf{k}} & v_{\mathbf{k}}\\
v_{-\mathbf{k}} & u_{-\mathbf{k}}
\end{bmatrix} \begin{bmatrix}
a_{\mathbf{k}} &\\
b_{-\mathbf{k}}^\dag
\end{bmatrix}
\end{equation} 

where,

\begin{align}
u_{{\bf k}}&=\sqrt{  \frac{1}{2}+ \frac{(J_0^{AB}-J_0^{AA}+J_{{\bf k}}^{AA})}{2\sqrt{(J_0^{AB}-J_0^{AA}+J_{{\bf k}}^{AA})^2-(J^{AB}_{{\bf k}})^2}}} \nonumber\\ 
v_{{\bf k}}&=sign(J^{AB}_{{\bf k}})\sqrt{u^2_{{\bf k}} - \frac{1}{2}}
\label{eq:u_kv_k}
\end{align}

\begin{align}
J^{AB}_{{\bf k}}&=\frac{1}{2}J_1(\cos(k_xa)+\cos(k_ya))\nonumber\\ 
J^{AA}_{{\bf k}}&=J^{AA}_{{\bf k}}=J_2(\cos(k_xa)\cos(k_ya))
\label{eq:JAA_JBB}
\end{align}

Here $a$ is the lattice constant. 
Hence,
\begin{equation}
\hat{H} = \sum_{{\bf k}} \epsilon_{\bf k} ( \alpha^\dag_{{\bf k}}\alpha_{{\bf k}}+\beta^\dag_{{\bf k}}\beta_{{\bf k}}) + \text{const} \label{eq:eigeb_value}
\end{equation}

\begin{equation}
\epsilon_{\bf k}= \sqrt{(J_0^{AB}-J_0^{AA}+J_{{\bf k}}^{AA})^2-(J_{{\bf k}}^{AB})^2}\label{eq:SMagnon}
\end{equation}   


\textit{The RIXS intensity:} The RIXS cross-section is given by Kramers-Heisenberg formalism (KH); $I_\text{RIXS} = \sum_{f}\vert \langle f| D_\text{out} \mathcal{O} D_\text{in}|g\rangle|^2 \delta (E_f-E_g-\omega)$. 
Here, $|g\rangle(|f\rangle)$ are the ground (final) states from the Hamiltonian $H_0$ with energy $E_g~(E_f)$, and $\omega$ is the energy loss. $D_\text{in(out)}$, is the dipole operator, and $\mathcal{O}$ accounts for the evolution of the system given in the presence of the core hole. We refer to recent literature for a detailed exposition of the simplification of the cross-section (See Ref.~\cite{UKumar2022}).  Following the literature~\cite{UKumar2022, PhysRevX.6.021020}, we employ ultra-short core-hole lifetime approximation. This introduces a broadening factor ($\Gamma$), which is the inverse of the core-hole lifetime. In the Appendix~\ref{ucl}, we outline that a perturbation in ($J/\Gamma$)  leads to two distinct contributions to the RIXS intensity, namely, the non-spin conserving (NSC, $\Delta S = 1$) and spin conserving  (SC, $\Delta S = 0$) channels. In what follows, we set $\hbar=1$. The RIXS intensity $I_\text{RIXS} \propto \sum_{l} S_{l}^\text{NSC}(\mathbf{q},\omega)+
\sum_{l}S_{l}^\text{SC}(\mathbf{q},\omega)$. Here $\mathbf{q}~(=\mathbf{k}_\text{out}-\mathbf{k}_\text{in}$) is the momentum transfer to the lattice.
The proportionality constant involves polarization dependent matrix elements arising out of the dipole operators and are different for the NSC and SC channels (see Ref.~\cite{PhysRevLett.103.117003}). In the Appendix~\ref{ucl}, we provide the SC and NSC UCL-expansion based on the contributions to the RIXS intensity. We explore both channels using the following response functions in our work:

i) {\sl Non-spin conserving channel:---}  The RIXS intensity can be decomposed into a sum of  terms  of $O(\frac{1}{\Gamma^{2l}})$ with the $l^{th}$ order contribution to the intensity given by
\begin{equation*}
\begin{split}
S_l^\text{NSC} (\mathbf{q}, \omega) = \frac{1}{\Gamma^{2l+2}}&\sum_{f}\left\vert \langle f|\frac{1}{\sqrt{N}}\sum_{i} e^{i {\bf q}\cdot {\bf R}_i} O_{i,l}^\mathrm{NSC} |g\rangle \right\vert^2 \\ 
& \times\delta\left(E_{f} - E_{g} - \omega\right),
\end{split}
\end{equation*}
From the Eq\ref{NSC-L} of Appendix~\ref{ucl}, we have: i) $O_{i,0}^\mathrm{NSC} = S_{i}^x $   for  $l=0$, ii) $O_{i,1}^\mathrm{NSC} = \sum_{j\in~\text{NN +NNN of}~i}J_{i,j}S_{i}^x \textbf{S}_i\cdot \textbf{S}_{j}$ , where $i$ is the location where the core-hole is created and,  iii) $O_{i,2}^\mathrm{NSC} = \sum_{j\neq k; j,k\in~\text{NN of}~i}J_{i,j}J_{i,k}S_{i}^x \textbf{S}_{j}\cdot \textbf{S}_{k}$ for $l=2$. 
We pictorially represent the summed over bonds around $i$ considered in $O_{i,1}^\mathrm{NSC}$ and $O_{i,2}^\mathrm{NSC}$ later in the paper.  

b) {\sl Spin conserving channel:---}  The $l^{th}$ order contribution to the RIXS intensity in the spin-conserving channel is: 
\begin{equation*}
\begin{split}
S_l^\text{SC} (\textbf{q}, \omega) = \frac{1}{\Gamma^{2l+2}}&\sum_{f}\left\vert \langle f|\frac{1}{\sqrt{N}}\sum_{i} e^{i {\bf q}\cdot {\bf R}_i} O_{i,l}^\mathrm{SC} |g\rangle \right\vert^2 \\ 
& \times\delta\left(E_{f} - E_{g} - \omega\right),
\end{split}
\end{equation*}
From the Eq\ref{SC-L} of Appendix~\ref{ucl}, we have: i)  $O_{i,0}^\mathrm{SC} = n_{i} $, ii) $O_{i,1}^\mathrm{SC} = \sum_{j\in~\text{NN +NNN of}~i}J_{i,j}\textbf{S}_i\cdot \textbf{S}_{j}$, and iii)  $O_{i,2}^\mathrm{SC}=\sum_{j\neq k; j,k\in~\text{NN of}~i}J_{i,j} J_{i,j}J_{i,k}(\mathbf{S}_i\cdot  \mathbf{S}_j) (\mathbf{S}_i\cdot  \mathbf{S}_k)$ as above. Notice that the $l=0$ order term does not lead to magnetic excitations and contributes to only the elastic scattering in the RIXS cross-section.  

\section{Results and Discussions}\label{sec:ResultsAndDiscussion}
We investigate the spin Hamiltonian given by Eq.~\ref{eq:hamiltonian0}  and limit ourselves to the linear spin wave theory approach. Within the UCL approximation of the KH formalism, the RIXS spectra can be mapped to a set of correlation functions in the SC and NSC channels, as presented above. We here report the responses till the second-order corrections of the UCL approximation. 
In this work, we set the work in the regime, $|J_2| < J_1$. We consider both positive (negative) $J_2$ suppressing (promoting) the overall $J_2=0$ AFM order. 

\begin{figure}[t]
	\centering
	\includegraphics[width=1\linewidth]{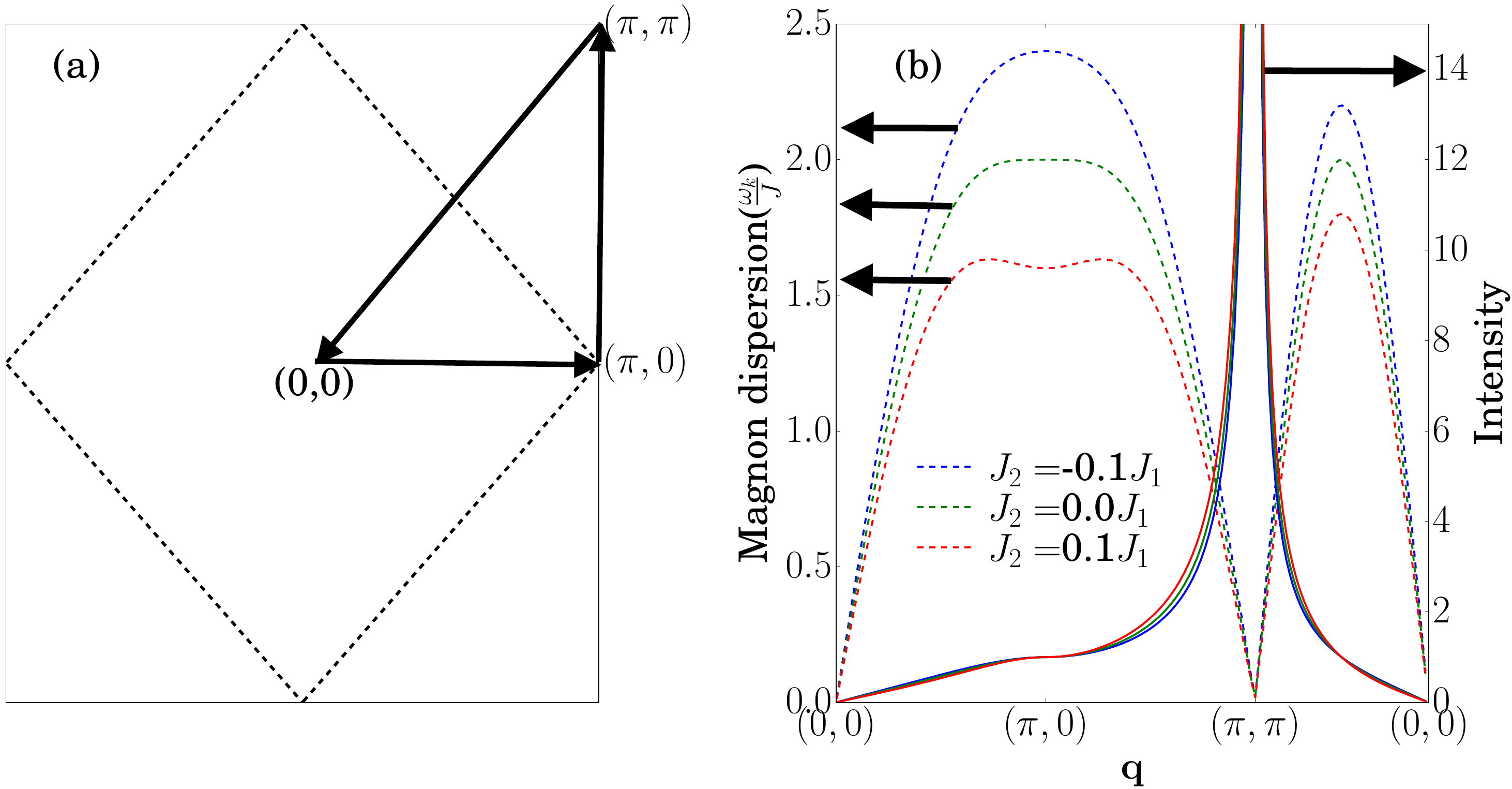}
	\caption{Dashed lines in panel (a) show the magnetic Brillouin zone boundary. Panel (b) shows the magnon dispersion (dashed line) and the momentum dependence  of one magnon integrated RIXS intensity  (solid lines) for the 2D extended Heisenberg model. The momentum is taken along the arrows, shown in panel (a).}
	\label{fig:Sqw}
\end{figure}

We present the Brillouin zone and momentum path for the 2D lattice in Fig.~\ref{fig:Sqw}(a) explored for the ease of the readers. We plot our results in $\Gamma (0, 0)-X (\pi, 0)-M (\pi, \pi)-\Gamma (0, 0)$ as highlighted in the figure unless otherwise stated. 

As there is limited discussions on the higher-order corrections in the NSC channel, we start by presenting our results in this channel. Later, we report the results in the relatively SC channel, extending it with long-range interactions in the correlation functions. 

\subsection{Non spin conserving channel}
In this section, we present results till the second-order correction in the UCL approximations. One peculiar feature of this channel is that there are an odd number of spin flips due to additional spin-flip mediated by the core-hole orbital with large spin-orbit coupling. Therefore, this channel is usually forbidden at the $K$-edge. \\

{\textit{Zeroth order}:---} At the zeroth order in the UCL expansion, the NSC channel allows for a single spin-flip mediated by the core-hole orbital. The scattering operator is, therefore, given by
\begin{equation}
O_{\mathbf{q}, 0}^\text{NSC}=\sum_{i}e^{i\mathbf{q}\cdot\mathbf{r}_i}S_{i}^x \label{onemag}. 
\end{equation}

We have preferentially chosen the spin-flip along the plane (along $x$-direction, here). We have to do this as the bipartite lattice used for the evaluating ground state fixes a quantization axis. If we can write the ground state without this in the ground state, then one can consider either component of the operator for spin~\cite{UKumar2022}. 
Using the LSWT approach discussed above, the operator can be mapped to the AFM magnon basis and is given by 
\begin{equation}
O_{\mathbf{q}, 0}^\text{NSC} \approx \sqrt{\frac{N}{2}}\sum_{i}(u_{\mathbf{q}}-v_{\mathbf{q}})(\alpha_{\mathbf{q}}+\beta_{\mathbf{q}}+\alpha_{\mathbf{-q}}^\dag+\beta_{\mathbf{-q}}^\dag) \label{eq:onemag_q}
\end{equation}
Notice that the operator can create a magnon on the vacuum, leading to a single magnon scattering. 


\begin{figure*}[ht]
	\centering
	\includegraphics[width=0.9\linewidth]{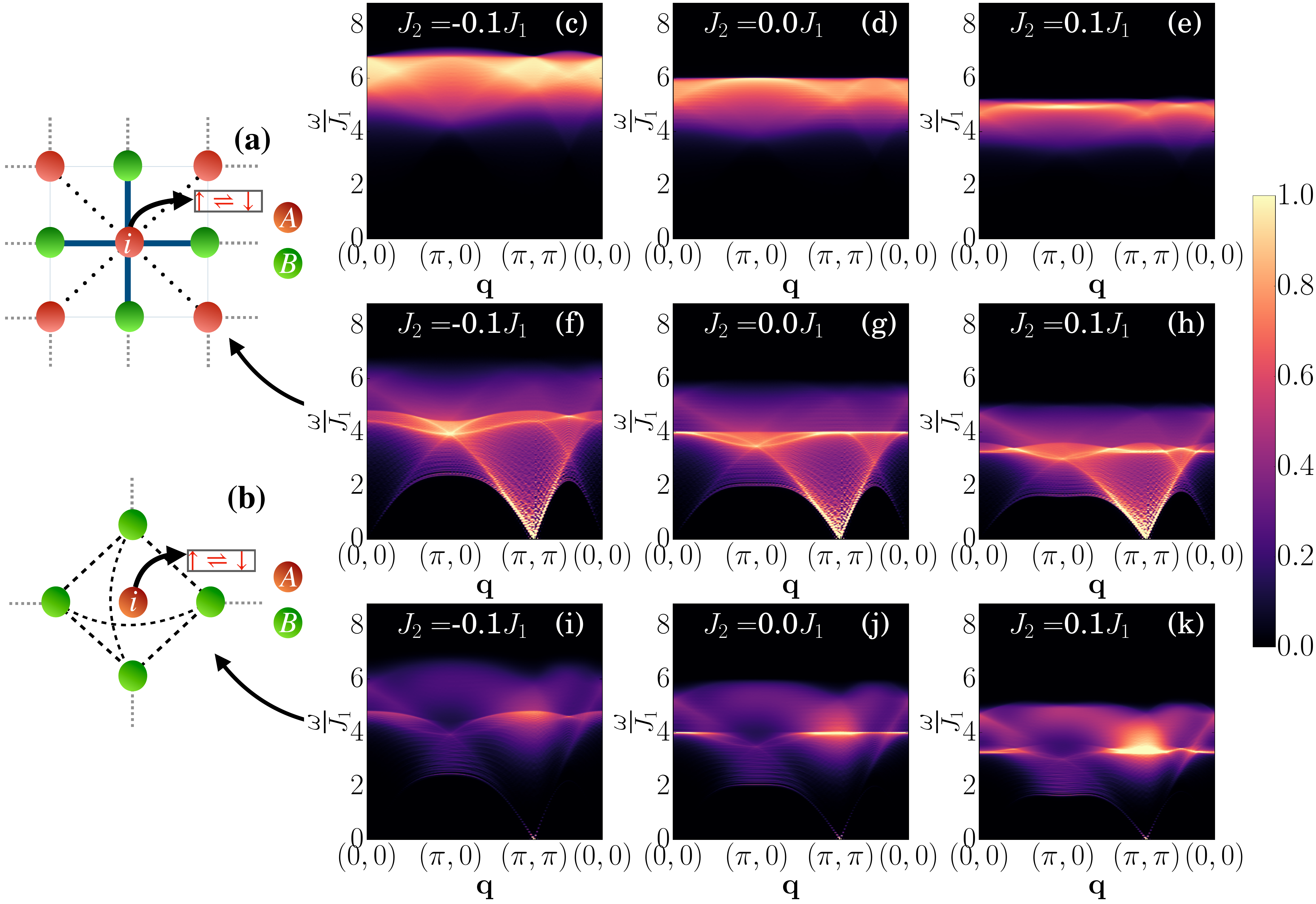}
	\caption{RIXS spectra in the non spin conserving channel (NSC).  Panel (a) and (b) show the schematic of first-order and second-order terms' contributions to the RIXS spectra, respectively, in the NSC channels corresponding to Eq \ref{eq:threemag_nsc} and Eq \ref{eq:threemag}. (c), (d) and (e) show the three-magnon DOS for the extended Heisenberg antiferromagnet for $J_2=-0.1 J_1, 0, 0.1 J_1$, respectively. 
 (f), (g) and (h) show the first-order correction of spectra for $J_2=-0.1 J_1, 0, 0.1 J_1$, respectively. The intensity in panels (f), (g) and (h) are multiplied by 100 for better visualization.
 Panels (i), (j), and (k) show the second-order correction of RIXS spectra for $J_2=-0.1 J_1, 0, 0.1 J_1$. 
 }
	\label{fig:threemag}
\end{figure*}

Fig.~\ref{fig:Sqw}(b) shows the \textit{zeroth} order response for 2D AFM lattice with NNN couplings. This is also equivalent to the dynamical spin structure factor, $S(q,\omega)$, observed in INS. We plot the magnon dispersion, evaluated using LSWT and the integrated intensity for the $S(q,\omega)$ response. We notice hardening of magnon dispersion for negative $J_2$, prominently at $(\pi, 0)$ and $(\pi/2, \pi/2)$. In particular, the hardening at $(\pi,0)$ is relatively larger compared to $(\pi/2, \pi/2)$. Conversely, there is softening for positive $J_2$. The signatures of these are also seen in the sharpening of the intensity around $(\pi,\pi)$ for $J_2<0$ compared to $J_2=0$. Our negative $J_2$ results are consistent literature ~\cite{PhysRevLett.103.117003}. Also positive $J_2$ results clearly weaken the AFM order. Such studies by varying the relative magnitude of $J_2/J_1$ has been used to explain certain aspects of the RIXS features of 2D cuprates~\cite{Peng2017} in the recent past.\\

\subsection*{Non-spin conserving channel: Higher orders}
As mentioned earlier, the NSC channel involves an odd number of spin flips. The higher-order correction can be described as a product of the single spin-flip operator and double spin-flip operators.  In principle, single and bi-magnon excitations can be created both independently as well as combined.  
Fig.~\ref{fig:threemag} (a) and (b) depict the first and second-order NSC excitations as discussed below. Since these excitations depend on the three-magnon density of states (DOS), we first discuss them.

The three-magnon DOS $A_{3M}(\omega)$, is given by the convolution of three one-magnon DOS and is given by $A_{3M}(\omega)=\sum_{\omega^\prime,\omega^{\prime\prime}}A_{1M}(\omega-\omega^\prime-\omega^{\prime\prime})A_{1M}(\omega^\prime)A_{1M}(\omega^{\prime\prime}) $.
Here $A_{1M}(\omega)=\sum_\mathbf{k}\delta(\omega-\epsilon_\mathbf{k}$), the dispersion  $\epsilon_\mathbf{k}$ is given in Eq.~\ref{eq:SMagnon}.
The three-magnon DOS is shown in Fig. \ref{fig:threemag}(c),(d), and (e). They result from the convolution of three one-magnon DOS for $J_2=-0.1, 0.0$, and $0.1$, respectively. We notice hardening in both the energy of DOS as well as spectral weight of three-magnon for negative $J_2$, whereas softening for positive $J_2$. In addition, we observe that the three-magnon DOS shows weak $\mathbf{q}$ dependence compared to the other cases discussed above. 

{\textit{First order}:---} The first order correction in the UCL expansion of the NSC channel leads to correction with a prefactor 
at $O(J^2/\Gamma^4)$ (where $J$ is the typical magnetic-exchange $J_1$), as shown in Eq \ref{NSC-L} of Appendix~\ref{ucl}. The scattering operator is given by
\begin{equation}
O_{\mathbf{q}, 1}^\text{NSC}=\sum_{ i,j }e^{i\mathbf{q}\cdot\mathbf{r}i}J_{i,j}S^x_i(\mathbf{S}_{i}\cdot \mathbf{S}_{j}).
\label{eq:threemag_nsc}
\end{equation}

Here, the sum over $i,j$ run over the NN and NNN sites as in the Hamiltonian $H_0$. Notice that the operator involves a single spin flip along with double spin flips.  The schematic is shown in Fig \ref{fig:threemag} (a).  
Solid and dotted lines represent the NN and NNN terms, respectively. 
Keeping only the linear terms in LSWT, the operator (at A) can  be mapped to bipartite bosons as
\begin{equation}
\begin{split}
S^x_i \mathbf{S}_{i}\cdot \mathbf{S}_{j} &\approx \sqrt{2S}(a_i^\dagger + a_i^\pdag) \Big(  -S^2 \\ 
& + S \big(a_i^\pdag b_j^\pdag + a_i^\dagger b_j^\dagger+a_i^\dagger a_i^{\pdag}+b_j^\dagger b_j^{\pdag} \big)  \Big)
\end{split}
\end{equation}

We can rewrite the operator as; $O_{\mathbf{q}, 1}^\text{NSC}= O_{\mathbf{q},1'}^\text{NSC}+O_{\mathbf{q},1''}^\text{NSC}$. The first term, $O_{\mathbf{q},1'}^\text{NSC}$, consists of only a single bosonic operator and can be mapped to a single magnon operator. The second term,  $O_{\mathbf{q},1''}^\text{NSC}$, consists of three bosonic operators and can be mapped to three-magnons which are given by
\begin{align}
    O_{\mathbf{q}, 1''}^\text{NSC}=\frac{S^{3/2}}{\sqrt{2N}}\sum_{\mathbf{k},\mathbf{p}}[f_0(\mathbf{k},\mathbf{p},\mathbf{q})& \alpha_{-\mathbf{p}}^\dag \beta_{-\mathbf{k}}^\dag\alpha_{\mathbf{k}+\mathbf{p}+\mathbf{q}}^\dag \nonumber\\+f_1&(\mathbf{k},\mathbf{p},\mathbf{q})\beta_{-\mathbf{p}}^\dag \beta_{-\mathbf{k}}^\dag\alpha_{\mathbf{k}+\mathbf{p}+\mathbf{q}}^\dag]
\label{eq:three_nsc_oq}
\end{align}
The detailed expressions for $f_0(\mathbf{k},\mathbf{p},\mathbf{q})$ and $f_1(\mathbf{k},\mathbf{p},\mathbf{q})$ term is presented in the Appendix \ref{Calculation details}.
In Fig.~\ref{fig:threemag}, panels (f), (g), and (h) show the response for the first order correction given from the first order corrections for $J_2 = -0.1J_1, J_1, 0.1J_1$, respectively. This has a weak one-magnon feature and three-magnon. The three-magnon ($J_2=0$) has three primary features: i) a dispersing band up to 4J with a bandwidth of $4J_1$, ii) a band localized at $4J_1$, and iii) a band that appears to disperse up to $5J_1$. 
Panels (f) and (h) show the results finite $J_2$, and highlight that these features are pushed to higher and lower energy for $J_2<0$ and $J_2>0$, akin to the single magnon case discussed previously.


{\textit{Second Order}:---} The second order correction consists of a single spin-flip along with a square of the double spin-flips and contributes at $\sim O(J^4/\Gamma^6)$. 
The scattering operator is given by
\begin{equation}
O_{\mathbf{q}, 2}^\text{NSC}=\sum_{\langle i,j,k \rangle}e^{i\mathbf{q}\cdot\mathbf{r}i}J_{ij}J_{ik}S^x_i(\mathbf{S}_{i}\cdot \mathbf{S}_{j})(\mathbf{S}_{i}\cdot \mathbf{S}_{k}).
\label{eq:secondorder_nsc}
\end{equation}
The double spin flip part can be rewritten as~\cite{PhysRevB.77.134428, UKumar2022}
\begin{align}
&J_{i,j}J_{i,k}(\mathbf{S}_{i}\cdot \mathbf{S}_{j})(\mathbf{S}_{i}\cdot \mathbf{S}_{k})\nonumber\\
&\approx -\frac{1}{2}J_{i,j}^2(\mathbf{S}_{i}\cdot \mathbf{S}_{j})+ \frac{1}{4}J_{i,j}J_{i,k}(\mathbf{S}_{j}\cdot \mathbf{S}_{k})
 \nonumber\\
\label{eq:twomag_cor1}
\end{align}
The four-spin can effectively be mapped to two two-spin, with NN interactions and longer-range interactions. 
The first term has the spin-operator indices as $i,j$ run over the NN. Since this term has the same form as that in the operator in first order except for a different prefactor, it gives rise to no new features. We thus focus only on the second term of Eq\ref{eq:twomag_cor1}. The second term in the expression consists of long-range double spin-flips, $j$ and $k$ indices, which are NN of the core-hole site-i. The schematic for this is shown in Fig.~\ref{fig:threemag}(b)



 \begin{figure*}[ht]
 \centering
 \includegraphics[width=0.9\linewidth]{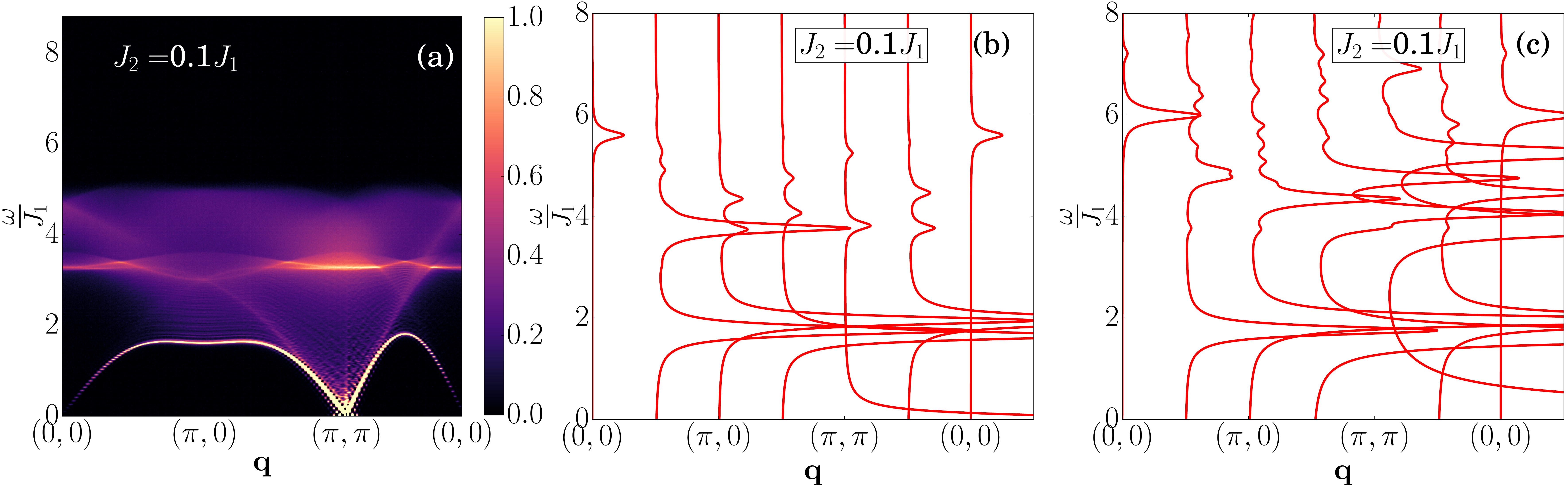}
\caption{Comparison of LSWT with the small cluster (4$\times$4 lattice) Exact Diagonalization results in the NSC channel. (a) shows the total contribution to the NSC channel up to the second-order in the UCL expansion evaluated using LSWT. Here $\Gamma$ is set equal to 5$J_1$. (b) and (c) show the ED results of the first and second-order contributions, respectively, for $J_2=0.1J_1$. The overall intensity in (b) has been scaled by a factor of 100 for better visibility.
}
  \label{fig:threemag_nsc-j2}
\end{figure*}

The new correlation in this order, therefore, is given by
\begin{equation}
O_{\mathbf{q}, 2'}^\text{NSC}=\sum_{i,j,k}e^{i\mathbf{q}\cdot\mathbf{r}_i}J_{i,j} J_{i,k} S^x_i(\mathbf{S_{j}}\cdot \mathbf{S}_{k})
\label{eq:threemag}
\end{equation}
As discussed for the first order, the HP transformation for this also yields two terms
\begin{equation}
\begin{split}
S^x_i \mathbf{S}_{j}\cdot \mathbf{S}_{k} &\approx \sqrt{2S}(a_i^\dagger + a_i^\pdag) \Big(  -S^2 \\ 
& + S \big(a_j^\pdag b_k^\pdag + a_j^\dagger b_k^\dagger+a_j^\dagger a_j^{\pdag}+b_k^\dagger b_k^{\pdag} \big)  \Big)
\end{split}
\end{equation}

Again, the operator can be rewritten as; $O_{\mathbf{q},2'}^\text{NSC}=O_{\mathbf{q},2'}^{\text{NSC},1}+O_{\mathbf{q},2'}^{\text{NSC},2}$. The first term, $O_{\mathbf{q},2'}^{\text{NSC},1}$, turns out to contribute to only a single spin-flip excitations akin to zeroth order. The second term has a distinct form and can contribute to multi-particle excitations. After Bogoliubov transformation, the second term is given by
\begin{align}
O_{\mathbf{q},2'}^\text{NSC,2} &\approx \frac{1}{\sqrt{2N}}\sum_{\mathbf{k},\mathbf{p}}f(\mathbf{k},\mathbf{p},\mathbf{q})[ (v_{\mathbf{p}} u_{\mathbf{k}+\mathbf{q}-\mathbf{p}}v_{\mathbf{k}} - u_{\mathbf{p}} v_{\mathbf{k}+\mathbf{q}-\mathbf{p}} \nonumber\\ & u_{\mathbf{k}}) \alpha_{\mathbf{p}}^\dag \alpha_{\mathbf{k}+\mathbf{q}-\mathbf{p}}^\dag\beta_{-\mathbf{k}}^\dag + 
(v_{\mathbf{p}} v_{\mathbf{k}+\mathbf{q}-\mathbf{p}}u_{\mathbf{k}} -  u_{\mathbf{p}} u_{\mathbf{k}+\mathbf{q}-\mathbf{p}}v_{\mathbf{k}})\nonumber\\ & \hspace{4.9cm}\times \beta_{\mathbf{p}}^\dag \alpha_{\mathbf{k}+\mathbf{q}-\mathbf{p}}^\dag\beta_{-\mathbf{k}}^\dag]
\label{eq:threemag_1}
\end{align}
where $f(\mathbf{k},\mathbf{p},\mathbf{q})=(J_1)^2[-6 \{ \cos(q_x-p_x)+\cos(q_y-p_y) \} + 2 \{ \cos(2k_x+q_x-p_x)+\cos(2k_y+q_y-p_y) \}+4 \{ \cos(k_y) \cos(k_x+q_x-p_x)+\cos(k_x) \cos(k_y+q_y-p_y) \}]$\\



We have, therefore, mapped this order to three-magnon excitations in the AFM lattice akin to the first order but with a distinct form.  

In Fig. \ref{fig:threemag}, panels (i), (j), and (k) show the corresponding three-magnon response function for Eq \ref{eq:threemag}. The three-magnon spectral weight has a clear feature peak around $\omega = 4J_1$ and at $\mathbf{q} = 0$ for $J_2=0$, that remains finite for all $q$ values in the Brillouin zone. The signature of one-magnon excitation is clearly seen in the sharp band in (i) dispersing through the Brillouin zone whose energy spread agrees with the single-magnon spectrum in Fig. \ref{fig:Sqw}(b). The intensity is, however, strongly suppressed. This feature has no contribution at $q=0$ due to vanishing one-magnon DOS. Similar features are also seen in Fig. \ref{fig:threemag} (j) and (k).
The continuum-like feature in (i) to (k) with finite spread across the full Brillouin zone arises from $O_{\mathbf{q}}^2$, which does not commute with $H_0$ at any $q$. 

Similar to the first order case, the responses with finite $J_2$ shown in panels (i) and (k) reveal softening (hardening) for $J_2>0$ ($J_2<0$). However, the intensity is stronger for $J_2>0$ and weakened for $J_2<0$. The latter is true because for $J_2<0$, the AFM order is strengthened, and thus spin-flip excitation arising from the AFM are more dominant than for $J_2<0$.  
We  conclude the LSWT discussion on the NSC channel by presenting in  Fig. \ref{fig:threemag_nsc-j2} (a) the consolidated contributions from the spin non-conserving channel up to second to the RIXS intensity.
For this plot, we set $J_1/\Gamma$ to 5 \cite{PhysRevB.77.134428}. We observe that relative to the one-magnon contribution, the higher-order contributions are significant.

\subsection*{Non-spin conserving channel: Exact Diagonalization}

We have also checked the consistency of the LSWT results against small cluster exact diagonalization. In Fig \ref{fig:threemag_nsc-j2} (b) and (c), we show the three-magnon susceptibility  on 4x4 lattice at first order and second order, respectively. In (b), we clearly see a small peak at ($\mathbf{q}=(0,0)$)  that disperses across the full Brillouin zone as we have concluded from the LSWT. We also see a clear energy gap between the low energy excitation, lying between $\omega=0$ to $\sim 2J_1$, and the higher energy excitations above $4J_1$.  Similarly, in (c), we observe that the low energy excitations are slightly pushed down below $2J_1$ but is separated from the higher energy excitations. By examining the basis state contributions to the various peaks over the Brillouin zone we find that the features below $\omega/J_1=2$, are primarily made out of one-flip basis states. This strongly suggests that the low energy LSWT spectra are indeed generated by single magnon excitations. Similarly, the high energy features are made up of multi-spin flip basis states.
However, due to severe size limitations the excitation energies are quite high, so the ED results simply provide a qualitative check of the contributions in the NSC channel. 
Thus the broad takeaway from the ED results is that the NSC channel produces a high energy excitation that has finite spectral weight over the entire Brillouin zone.



\subsection{Spin conserving channel}
Here, we discuss the contributions from the spin-conserving channel of the UCL approximation of the KH formalism. This channel has been discussed in the context of SC edges in RIXS. For example, two-magnon at the oxygen $K$-edge RIXS of cuprate was reported~\cite{PhysRevB.85.214527}. With the improved RIXS resolution, the SC channel can also be realized at Cu $L$-edge and has been explored in 1D chains~\cite{UKumar2022, PhysRevLett.112.147401}.  Therefore, we revisit this channel in the context of $L$-edge and provide results for $J_2 \neq 0$ for 2D lattice. In particular, the computed three-magnon excitations overlap with two-magnon DOS; these results help us rule out whether these NNN terms can contribute to the new phase allowed in the NSC channel.

\begin{figure*}[ht]
\centering
\includegraphics[width=0.9\linewidth]{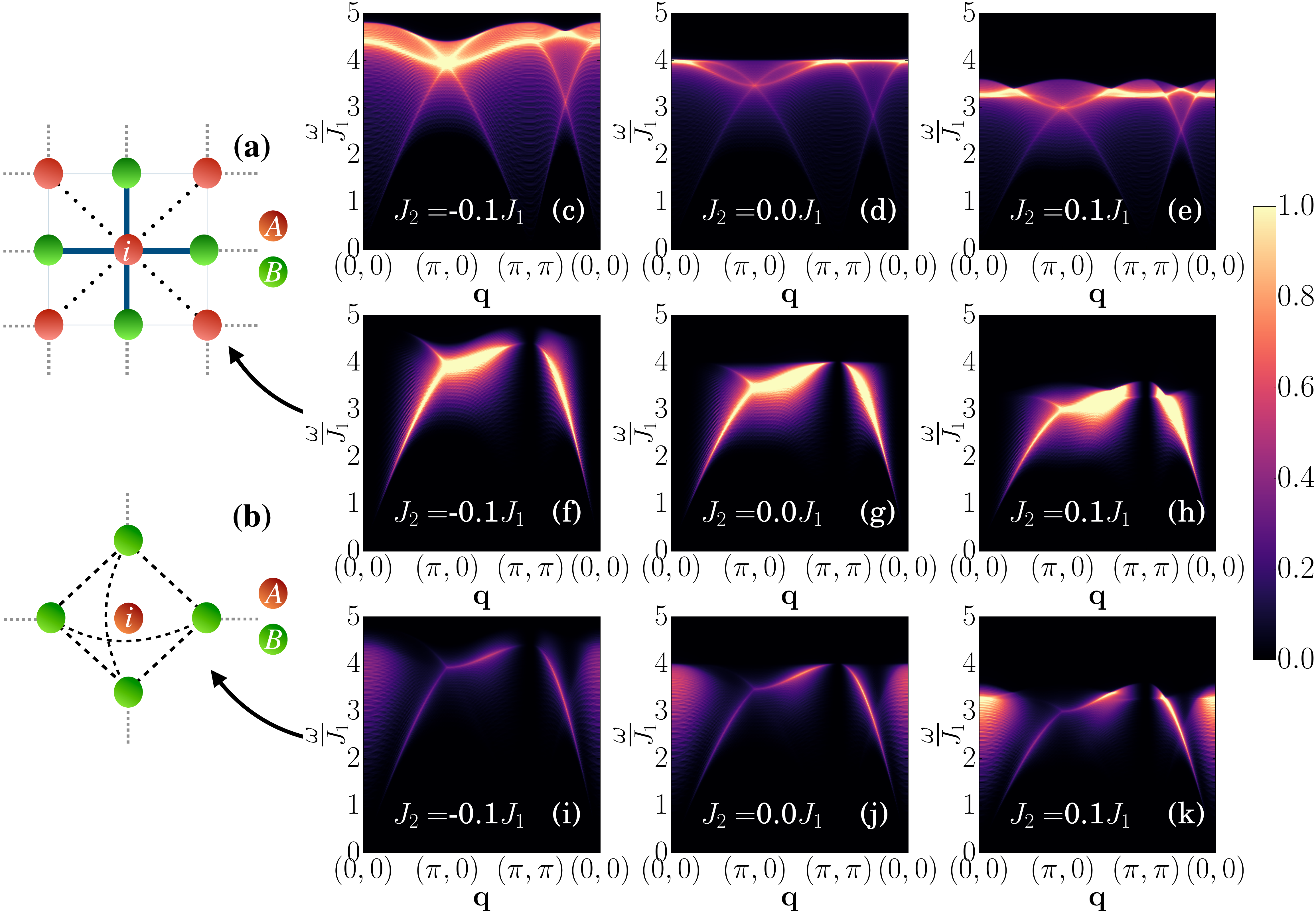}
\caption{ RIXS spectra in the spin conserving channel (SC).  Panel (a) and (b) show the schematic of first-order and second-order terms' contributions of SC channel to the RIXS spectra, respectively, corresponding to Eq \ref{eq:twomag_1} and Eq \ref{eq:twomag_cor2n}. (c), (d) and (e) show the two-magnon DOS for the extended Heisenberg antiferromagnet for $J_2=-0.1 J_1, 0, 0.1 J_1$, respectively. (f), (g) and (h) show the first-order correction of spectra for $J_2=-0.1 J_1, 0, 0.1 J_1$, respectively. Panels (i), (j), and (k) show the second-order correction of RIXS spectra for $J_2=-0.1 J_1, 0, 0.1 J_1$, respectively. 
}\label{fig:twomag1}
\end{figure*}
{\textit{Zeroth Order}:--} The SC channel does not allow any spin flip at the zeroth order. Therefore, magnetic excitation is forbidden and it contributes only to elastic scattering for the spin model. 

\subsection*{Spin conserving channel: Higher orders}
The higher order corrections in the SC channel is given by an even number of multi-spin-flips and can lead to magnetic excitations. Here, we investigate the dynamical correlation functions relevant to the two-magnon excitations in this channel. In this channel, for completeness, we also explore the correlation function with the NNN. This will further highlight why one needs the NSC channel to reproduce the novel features in the RIXS spectra of 2D AFMS. 

The leading corrections in this channel maps to the two-magnon.  The two-magnon DOS $A_{2M}(\omega)$ is given by the convolution of two one-magnon DOS, $A_{2M}(\omega)=\sum_{\omega^\prime}A_{1M}(\omega-\omega^\prime)A_{1M}(\omega^\prime)$, with $A_{1M}(\omega)=\sum_\mathbf{k}\delta(\omega-\epsilon_\mathbf{k}$), see Eq.~\ref{eq:SMagnon}
We show the two-magnon density of states (DOS) for $J_2=-0.1,0$ and $0.1$ in Fig.~\ref{fig:twomag1} (c)-(e).  


{\textit{First Order}:--} 
The first order correction in the UCL expansion of $O(J^2/\Gamma^4)$ is given by double spin-flips. The scattering operator for this is given by
\begin{equation}
O_{\mathbf{q}, 1}^\text{SC}=\sum_{i,j}e^{i\mathbf{q}\cdot\mathbf{r}_i}J_{i,j}\mathbf{S}_{i}\cdot \mathbf{S}_{j}
\label{eq:twomag_1}
\end{equation}
Here, the sum over $i,j$ run over the NN and NNN sites as in the Hamiltonian $H_0$. These NN (solid) and NNN (dashed) bonds are shown in Fig~\ref{fig:twomag1} (a). 

With the HP transformation, Eq \ref{eq:twomag_1} can be expressed as
\begin{align}
O_{\mathbf{q},1}^\text{SC}\approx S\sum_{\mathbf{k}\in BZ}[f_0(\mathbf{k},\mathbf{q}) 
\times(a_{\mathbf{k}-\mathbf{q}/2}^{\dag} a_{\mathbf{k}+\mathbf{q}/2} +   b_{\mathbf{k}-\mathbf{q}/2}^{\dag} b_{\mathbf{k}+\mathbf{q}/2}) \nonumber\\ +f_1(\mathbf{k},\mathbf{q})
(a_{\mathbf{k}-\mathbf{q}/2} b_{-\mathbf{k}-\mathbf{q}/2} +   a_{\mathbf{k}+\mathbf{q}/2}^{\dag} b_{-\mathbf{k}-\mathbf{q}/2}^{\dag} )]
\label{eq:twomag_ab1}
\end{align}
Finally the Bogoliubov transformation yields Eq \ref{eq:twomag_1} in the form

\begin{align}
O_{\mathbf{q},1}^\text{SC} \approx S\sum_{\mathbf{k}\in BZ}[-f_0(\mathbf{k},\mathbf{q}&) 
(u_{\mathbf{k}+\mathbf{q}/2} v_{\mathbf{k}-\mathbf{q}/2} +   u_{\mathbf{k}-\mathbf{q}/2} v_{\mathbf{k}+\mathbf{q}/2})\nonumber\\  +f_1(\mathbf{k},\mathbf{q})&
(u_{\mathbf{k}+\mathbf{q}/2} u_{\mathbf{k}-\mathbf{q}/2} +   v_{\mathbf{k}+\mathbf{q}/2} v_{\mathbf{k}-\mathbf{q}/2}^{\dag} )]\nonumber\\
(\alpha_{\mathbf{k}+\mathbf{q}/2}&\beta_{-\mathbf{k}+\mathbf{q}/2}+
\alpha_{\mathbf{k}-\mathbf{q}/2}^{\dag}\beta_{-\mathbf{k}-\mathbf{q}/2}^{\dag})
\label{eq:twomag_o}
\end{align}

where $f_0(\mathbf{k},\mathbf{q})=(J_{\mathbf{k}+\mathbf{q}/2}^{AA}  +J_{\mathbf{k}-\mathbf{q}/2}^{AA}+J_{0}^{AA}+J_{\mathbf{q}}^{AA}+J_{0}^{AB}+J_{\mathbf{q}}^{AB})$  and $f_1(\mathbf{k},\mathbf{q})=(J_{\mathbf{k}+\mathbf{q}/2}^{AB}+J_{\mathbf{k}-\mathbf{q}/2}^{AB})$.

The first order scattering operator in UCL approximation within the LSWT can be mapped to the two-magnon as discussed above.

 
In Fig.\ref{fig:twomag1}, panels (f), (g), and (h) show the response for the first-order correction in this channel.
A number of striking features can be observed. The spectral weight vanishes at $\mathbf{q}=(0,0)$ and $\mathbf{q}=(\pi,\pi)$, which follows from Eq:\ref{eq:twomag_ab1}. More intuitively, since $[H_0,O_{\mathbf{q}=(0,0)}]=0$, they have the same eigenbasis and hence the matrix element $\langle f|O_{\mathbf{q}=(0,0)}|g\rangle$ is zero.
Thus, the intensity vanishes at the zero transferred momentum, in agreement with the experimental observation \cite{PhysRevLett.100.097001}. Because of the antiferromagnetic ground state, the RIXS intensity always vanishes at  $\mathbf{q}=(\pi,\pi)$. The $\mathbf{q}=(\pi,\pi)$ is the reciprocal magnetic lattice vector for antiferromagnetic order. Setting $\mathbf{q}=(\pi,\pi)$ in the Eq\ref{eq:twomag_1} we find that   
$O_{\mathbf{q}=(\pi,\pi)}=\sum_{i\in A,j}\mathbf{S}_{i}\cdot \mathbf{S}_{j} -\sum_{i\in B,j}S_{i}\cdot S_{j}$, where $A$ and $B$ are two sub-lattice indexes. Initial state, $\ket{i}$, is symmetric under the interchange of sub lattice, whereas $O_{\mathbf{q}=(\pi,\pi)}$ is anti-symmetric under the interchange of sub lattice. Therefore $\bra{f}O_{\mathbf{q}=(\pi,\pi)}\ket{g}$ vanished identically.
We also notice an overall hardening in both two-magnon DOS as well as spectral weight for negative $J_2$ and softening for positive $J_2$. \\



{\textit{Second Order}:---}
The second order correction in the UCL expansion of $O(J^4/\Gamma^6)$ is given by
\begin{equation}
    O_{\mathbf{q}, 2}^\text{SC}=\sum_{i,j,k}e^{i\mathbf{q}\cdot\mathbf{r}_i} J_{i,j}J_{i,k}(\mathbf{S}_i\cdot  \mathbf{S}_j) (\mathbf{S}_i\cdot  \mathbf{S}_k)
    \label{eq:twomag_cor2n}
\end{equation}

Simplifying $J_{i,j}J_{i,k}(\mathbf{S}_i\cdot  \mathbf{S}_j) (\mathbf{S}_i\cdot  \mathbf{S}_k)$ similar to the NSC channel (see Eq.~\ref{eq:twomag_cor1}), there are two terms. One is the two-magnon term $(\mathbf{S}_i.\mathbf{S}_j)$ with a different prefactor, and a second term $ \frac{1}{4}J_{i,j}J_{i,k}(\mathbf{S}_{j}\cdot \mathbf{S}_{k})$. As in the NSC channel, the first term does not contribute to any new feature. The summation over the $j,k$ indices in the second term are shown in Fig.~\ref{fig:twomag1} (b). 
Following the approach discussed above, this term can be expressed as 

\begin{align}
&O_{\mathbf{q},2'}^\text{SC} \approx \sum_{\mathbf{k}}f(\mathbf{k},\mathbf{q})( u_{\mathbf{k}}v_{\mathbf{k}+\mathbf{q}} + u_{\mathbf{k}+\mathbf{q}} v_{\mathbf{k}}) \alpha_{\mathbf{k}}^\dag\beta_{-\mathbf{k}-\mathbf{q}}^\dag
\label{eq:twomag_cor2}
\end{align}
where $f(\mathbf{k},\mathbf{q})=(J_1)^2[-6\{\cos(q_x)+\cos(q_y)\} +2\{\cos(2k_x+q_x)+\cos(2k_y+q_y)\}+4\{\cos(k_y)\cos(k_x+q_x)+\cos(k_x)\cos(k_y+q_y)\}]$

In Fig.~\ref{fig:twomag1}, panels (i), (j), and (k) show the spectra for this response function for $J_2 = -0.1J_1, J_1$ and $0.1 J_1$ respectively. For $J_2=0$ shown in panel (j), we notice dispersive features exist over the energy window allowed by the two-magnon DOS. However, unlike the at the previous order contribution to the SC channel, we now observe clear weight at ${\bf q}=(0,0)$ spread between $\omega/J_1=3$ and 4. This weight however is limited between ${\bf q}=(0,0)$ and ${\bf q}=(\sim\pi/2,0)$. The other feature in the middle of the plot resembles the two-magnon features in the first order correction. Thus the main correction to the two-magnon spectra coming from second order are new spectral weights around ${\bf q}=(0,0)$. 

The two-magnon term discussed above commutes with $H_0$ at both the first and second order of the SC channel at $\mathbf{q}=(0, 0)$. However, the term $ \frac{1}{4}J_{i,j}J_{i,k}(\mathbf{S}_{j}\cdot \mathbf{S}_{k})$ does not commute with $H_0$ at $\mathbf{q}=0$ and gives rise to a finite spectral weight at second order.
The same features carry over to the cases with $J_2= \pm0.1 J_1$, with $J_2>0$ ($J_2<0$), making the features much more dominant (suppressed) as is expected. 

\section{Conclusion}\label{sec:Conclusion}
We have provided a comprehensive study of the  RIXS cross-section for 2D AFM in both the NSC and SC channels up to the second order corrections in the UCL expansion of the KH formalism relevant for Cu $L$-edge of 2D cuprates. We explore these corrections using LSWT. We report the observation of three-magnon excitations in the NSC channel. We find that three-magnon has finite weights in both the first and second-order corrections in the NSC channels, albeit the weights are larger in second-order correction.  These three-magnon excitations have a clear high energy feature with a quasi-flat band extending over the entire Brillouin zone, very distinct from the two-magnons reported in the literature earlier.
We further report that the LSWT results of NSC channel agree qualitatively with exact diagonalization results on a small cluster.

The Cu $L$-edge RIXS in 2D cuprates has a large contribution from the NSC channel.
We, therefore, compare our results with the Cu $L_3$-edge RIXS data of La$_2$CuO$_4$ reported in Ref.~\cite{PhysRevB.103.224427}. It shows a peculiar excitation at around $\omega= 350$ meV, which weakly disperses  over the Brillouin zone centered and extends upto at ${\bf q}=(0,0)$, in addition to the one-magnon and two magnon. 
For $J_1= 350$ meV, we find that the spread of the RIXS spectra qualitatively agrees with the RIXS experiment, and this feature can, therefore, be interpreted as three-magnon. Nevertheless, our calculations restricted to the LSWT predict higher energy for the observed three-magnon than the RIXS data. Quantitative agreement is not expected as our work lacks magnon-magnon interaction, which can lower the effective energies of these excitations.

We further report the SC channel of the RIXS and extend the correlation functions with long-range interaction relevant for Cu $L$-edge in cuprates. We find that the longer-range interaction in this channel does not qualitatively alter the features of the first and second-order corrections. This suggests that the higher-order corrections of the NSC channel are critical to reproduce the magnetic excitations in the new phase space observed at $L$-edge with improved resolution of RIXS.


Thus, our result is an important step in identifying the nature of multi-magnon excitations observed in the RIXS spectra of 2D antiferromagnets~\cite{PhysRevB.103.224427}. The relevance of higher order correction in the RIXS cross-section in the 2D AFMs opens up new pathways to explore the higher modes of magnetic excitations in quantum materials using RIXS.

\section{Acknowledgments}  
All computations were performed in the NOETHER, VIRGO, and KALINGA high performance clusters at NISER. A.M. would like to acknowledge  the MATRICS grant (Grant No. MTR/2022/000636)  from the Science and Engineering Research Board (SERB) for funding. 

\appendix*
 \setcounter{equation}{0}  
\begin{center}
	\textbf{APPENDIX}
\end{center}

\subsection{UCL expansion for Kramers-Heisenberg formalism}\label{ucl}

Here we provide the relevant expressions for the NSC and the SC scattering cross-sections. We refer the reader to recent literature for details ~\cite{UKumar2022, PhysRevX.6.021020}. 
At the $L$-edge, the dipole operator can allow for single spin-flip excitations leading to the following spin non-conserving contributions to the  RIXS cross-section at various orders of the inverse core-hole lifetime parameter ($\Gamma$) : 

\begin{widetext}
\begin{align}
I^\text{NSC}(\mathbf{q},\omega) & \propto \big( \frac{1}{\Gamma^2}\sum_f \Big|\langle f|\frac{1}{\sqrt{N}}\sum_i e^{i\mathbf{q}\cdot \mathbf{R}_{i}}S_i^{x}|g\rangle\Big|^2
+ \frac{1}{\Gamma^4}\sum_f \Big|\langle f|\frac{1}{\sqrt{N}}\sum_{i,j} e^{i\mathbf{q}\cdot \mathbf{R}_{i}}J_{i,j}S_i^{x}(\hat{S}_i\cdot  \hat{S}_j) |g\rangle\Big|^2 &\nonumber\\
&+\frac{1}{\Gamma^6}\sum_f \Big|\langle f|\frac{1}{\sqrt{N}}\sum_{i,j,k} e^{i\mathbf{q}\cdot \mathbf{R}_{i}}J_{i,j}J_{i,k}S_i^{x}(\hat{S}_i\cdot  \hat{S}_j)(\hat{S}_i\cdot  \hat{S}_k) |g\rangle\Big|^2 + \cdot \cdot\cdot\Big)\delta\left(E_f-E_g-\omega\right)= \sum_{l}S_{l}^{NSC}(\mathbf{q},\omega)
\label{NSC-L}
\end{align}
\end{widetext}

In the above the $O((1/\Gamma)^2)$  term is the single spin-flip spin excitation scattering and the $O((1/\Gamma)^4)$ term is a combination single spin-flip at the site where the core-hole is created and a bi-magnon living either the $j$ and $k$ sites NN to $i$, or a bi-magnon involving $i$ and $j$, or $i$ and $k$ as discussed in the paper. 

For the spin-conserving channel, the contributions to the RIXS intensity are given by
\begin{widetext}
\begin{align}
I^\text{SC}(\mathbf{q},\omega) & \propto \big( \frac{1}{\Gamma^2}\sum_f \Big|\langle f|\frac{1}{\sqrt{N}}\sum_i e^{i\mathbf{q}\cdot \mathbf{R}_{i}}n_{i,\sigma}|g\rangle\Big|^2
+ \frac{1}{\Gamma^4}\sum_f \Big|\langle f|\frac{1}{\sqrt{N}}\sum_{i,j} e^{i\mathbf{q}\cdot \mathbf{R}_{i}}J_{i,j}\hat{S}_i\cdot  \hat{S}_j |g\rangle\Big|^2 &\nonumber\\
&+ \frac{1}{\Gamma^6}\sum_f \Big|\langle f|\frac{1}{\sqrt{N}}\sum_{i,j,k} e^{i\mathbf{q}\cdot \mathbf{R}_{i}}J_{i,j}J_{i,k}(\hat{S}_i\cdot  \hat{S}_j) (\hat{S}_i\cdot  \hat{S}_k)|g\rangle\Big|^2 + \cdot \cdot \cdot\textbf{}\Big)\delta\left(E_f-E_g-\omega\right)=\sum_{l} S_{l}^\text{SC}(\mathbf{q},\omega).
\label{SC-L}
\end{align}
\end{widetext}

We note that the $O((1/\Gamma)^2)$ term does not contribute to spin excitations. The higher order terms lead to magnetic excitations. The bi-magnon excitations from these higher orders are calculated and discussed in the paper.
\subsection{Calculation details}\label{Calculation details}
The detailed expressions for $f_0(\mathbf{k},\mathbf{p},\mathbf{q})$ and $f_1(\mathbf{k},\mathbf{p},\mathbf{q})$ used in  Eq:\ref{eq:three_nsc_oq}  in the paper are provided below, with $J^{NNN}_{\mathbf{k},\mathbf{p},\mathbf{q}}=J^{AA}_{\mathbf{k}+\mathbf{p}+\mathbf{q}}+J^{AA}_{\mathbf{k}}-J^{AA}_{0}-J^{AA}_{\mathbf{p}+\mathbf{q}}$:
\begin{widetext}
\begin{align}
f_0(\mathbf{k},\mathbf{p},\mathbf{q})=&-\{((J_{\mathbf{p}+\mathbf{q}}^{AB}+J^{NNN}_{\mathbf{k},\mathbf{p},\mathbf{q}})u_{\mathbf{p}}-((J_{0}^{AB}+J^{NNN}_{\mathbf{k},\mathbf{p},\mathbf{q}})v_{\mathbf{p}})u_{\mathbf{k}}v_{\mathbf{k}+\mathbf{p}+\mathbf{q}}+((J_{0}^{AB}+J^{NNN}_{\mathbf{k},\mathbf{p},\mathbf{q}})u_{\mathbf{p}}-(J_{\mathbf{p}+\mathbf{q}}^{AB}+J^{NNN}_{\mathbf{k},\mathbf{p},\mathbf{q}}) v_{\mathbf{p}})\nonumber\\
&v_{\mathbf{k}}u_{\mathbf{k}+\mathbf{p}+\mathbf{q}}\}+(J_{\mathbf{k}+\mathbf{p}+\mathbf{q}}^{AB}u_{\mathbf{p}}-J_{\mathbf{k}}^{AB}v_{\mathbf{p}})v_{\mathbf{k}+\mathbf{p}+\mathbf{q}}v_{\mathbf{k}}+(J_{\mathbf{k}}^{AB}u_{\mathbf{p}}-
J_{\mathbf{k}+\mathbf{p}+\mathbf{q}}^{AB}v_{\mathbf{p}})u_{\mathbf{k}+\mathbf{p}+\mathbf{q}}u_{\mathbf{k}}  
\end{align}
\begin{align}
f_1(\mathbf{k},\mathbf{p},\mathbf{q})=&-\{((J_{0}^{AB}+J^{NNN}_{\mathbf{k},\mathbf{p},\mathbf{q}})u_{\mathbf{p}}-(J_{\mathbf{p}+\mathbf{q}}^{AB}+J^{NNN}_{\mathbf{k},\mathbf{p},\mathbf{q}})v_{\mathbf{p}})u_{\mathbf{k}}v_{\mathbf{k}+\mathbf{p}+\mathbf{q}}+((J_{\mathbf{p}+\mathbf{q}}^{AB}+J^{NNN}_{\mathbf{k},\mathbf{p},\mathbf{q}})u_{\mathbf{p}}-(J_{0}^{AB}+J^{NNN}_{\mathbf{k},\mathbf{p},\mathbf{q}})v_{\mathbf{p}})\nonumber \\ 
&v_{\mathbf{k}}u_{\mathbf{k}+\mathbf{p}+\mathbf{q}}\} +(J_{\mathbf{k}}^{AB}u_{\mathbf{p}}-J_{\mathbf{k}+\mathbf{p}+\mathbf{q}}^{AB}v_{\mathbf{p}})v_{\mathbf{k}+\mathbf{p}+\mathbf{q}}v_{\mathbf{k}}+(J_{\mathbf{k}+\mathbf{p}+\mathbf{q}}^{AB}u_{\mathbf{p}}-J_{\mathbf{k}}^{AB}v_{\mathbf{p}})u_{\mathbf{k}+\mathbf{p}+\mathbf{q}}u_{\mathbf{k}}
\end{align}
\end{widetext}
\bibliography{bibfile}

\end{document}